# Recent progress in molecular simulation methods for drug binding kinetics

Ariane Nunes-Alves[a,b+], Daria B. Kokh[a+], Rebecca C. Wade[a,b,c*]

[a]Molecular and Cellular Modeling Group, Heidelberg Institute for Theoretical Studies, Schloss-Wolfsbrunnenweg 35, 69118 Heidelberg, Germany
[b]Center for Molecular Biology (ZMBH), DKFZ-ZMBH Alliance, Heidelberg University, Im Neuenheimer Feld 282, 69120 Heidelberg, Germany
[c]Interdisciplinary Center for Scientific Computing (IWR), Heidelberg University, Im Neuenheimer Feld 205, Heidelberg, Germany

ariane.nunes-alves@h-its.org, daria.kokh@h-its.org, Rebecca.wade@h-its.org

*Corresponding author: Rebecca.wade@h-its.org, Tel: +49 6221 533247
+ These authors are equal first authors

**Abstract:**

Due to the contribution of drug-target binding kinetics to drug efficacy, there is a high level of interest in developing methods to predict drug-target binding kinetic parameters. During the review period, a wide range of enhanced sampling molecular dynamics simulation-based methods has been developed for computing drug-target binding kinetics and studying binding and unbinding mechanisms. Here, we assess the performance of these methods considering two benchmark systems in detail: mutant T4 lysozyme-ligand complexes and a large set of N-HSP90-inhibitor complexes. The results indicate that some of the simulation methods can already be usefully applied in drug discovery or lead optimization programs but that further studies on more high-quality experimental benchmark datasets are necessary to improve and validate computational methods.

## Introduction

The realization that drug binding kinetic parameters can be key determinants of drug efficacy in non-equilibrium *in vivo* conditions [1,2*], has led to interest in developing methods to compute ligand-receptor association ($k_{on}$) and dissociation ($k_{off}$) rate constants. Two years ago, we reviewed emerging computational methods [3] and noted that most of these had been applied to and validated on very small datasets, sometimes only one protein-ligand complex. Over the last two years, the period of focus of this review, the rate of publication of methods to study binding kinetics has burgeoned. Some of these studies continue to be based on small datasets and have focused on exploring mechanistic aspects of binding kinetics by molecular simulation approaches. These studies are exemplified by eleven publications on a T4 lysozyme (T4L) mutant with a buried cavity in which benzene derivatives bind. We compare the approaches taken and the results obtained in the next section.

The emergence of larger, systematically measured sets of kinetic data and structures of protein-ligand complexes provides a basis for developing computational methods suitable for deriving quantitative structure-kinetics relationships (QSKRs) and for their application in the lead optimization stage of drug discovery. Some of these studies aim at employing statistical machine learning or chemometric approaches to exploit structural information on protein-ligand complexes to derive QSKRs. However, the majority of the methods published in the last two years employ molecular dynamics (MD) simulation combined with enhanced sampling techniques. The most well-studied of these datasets is that for heat shock protein 90 (N-HSP90) inhibitors. Below, we compare the studies done on N-HSP90 inhibitors, the different methods, their performance and mechanistic insights.

By focusing on two selected protein targets, we aim to provide a practically relevant assessment of the current state of the field of computing binding kinetics, mechanisms and QSKRs. In this brief article, it is not possible to describe in detail all the relevant methods or the systems to which such methods have been applied. We therefore refer the reader to several recent reviews [2,4*–6] as well as to the on-line toolbox, KBbox [7] (kbbox.h-its.org), which contains descriptions of the different methods, along with some tutorials and guidance on usage.

## Exploring ligand pathways and associated kinetic rates: the L99A T4 lysozyme mutant

The L99A mutant of T4L has an engineered, buried cavity that can accommodate benzene derivatives (see **Figure 1**). Although it is not a drug target, T4L has been used as a model system to study protein-ligand binding, due to the small size and rigidity of the ligands and their fast binding and unbinding rates (with residence times, $\tau$ ($=1/k_{off}$), on the millisecond timescale for benzene and indole [8]). However, understanding T4L-ligand binding processes poses challenges for computational studies such as sufficiently sampling the protein conformational changes necessary to make the buried cavity accessible to ligands. The question of how many paths are available for ligand binding to and unbinding from T4L has been addressed in several recent studies. Complete sampling of ligand paths is important for calculating (un)binding rates. Moreover, knowledge of ligand paths and the associated (un)binding mechanisms can guide the design of ligands or proteins with altered binding kinetics. Ligand passage to and from binding sites cannot be directly observed experimentally and thus simulation methods have an important role in allowing such paths to be identified, visualized and quantitatively understood.

*De novo detection of tunnels and enhanced sampling of binding paths*
While binding of a ligand is often considered to occur by one pathway, computational studies of T4L show that multiple pathways exist and the number of paths identified varies between different studies, see **Figure 1**. The highest number of paths, eight, was identified by Capelli *et al.* using well-tempered metadynamics [9].

In other studies, fewer paths were identified. This inconsistency may in part be due to the different definitions of the pathways. For example, Rydzewski and Valsson sampled five paths for benzene unbinding from T4L (CF, CD, FGH, HJ, DG) [10] and 4-5 different unbinding paths for various benzene derivatives [11]. However, the authors noted [10] that four of the paths identified by Capelli *et al.* were subclasses of the path FGH, which had the widest tunnel width and is the one path observed in all studies. Moreover, this pathway was suggested as the most probable by Feher et al. [12]. However, in these studies, pathways were classified by visual inspection, which may lead to differences due to subjectivity.

Another reason for the observation of different numbers of pathways in different studies may be that some of the enhanced sampling procedures alter the energetic barriers to ligand binding and unbinding. Indeed, in an unbiased MD study, three paths for benzene binding (FGH, HJ, DG) were observed whereas, in the same study infrequently biased metadynamics (InMetaD), in which the transition state should not be affected by the addition of potential terms to enhance sampling along collective variables, two paths (FGH, HJ) for benzene unbinding were sampled [13].

A further reason for variations in the number of pathways is the ligand size and physicochemical properties. For instance, the binding of benzene and benzene derivatives to T4L was studied by Niitsu et al. using the generalized replica exchange with solute tempering (gREST) method together with a flat-bottomed potential to limit the sampling of ligand positions [14]. Two binding paths (DG and FGH) were identified with preferences for the paths depending on ligand size and indications that the binding of larger ligands through the preferred pathway (FGH) requires coupling with a weakly populated conformation of helix F. Sampling of this excited state was however identified in accelerated molecular dynamics (aMD) simulations as responsible for allowing benzene to exit through path FGH [12]. Indeed, nuclear magnetic resonance experiments on T4L [15,16] have shown a conformational exchange involving motion of helix F that occurs between the highly populated (crystallographic) structure and a weakly populated state on the millisecond timescale. This is the same timescale as that for benzene unbinding, and therefore the experiments and simulations together suggest that the ligand unbinding rate may be governed by the timescale of the accompanying protein motions.

*Computation of binding kinetic rates*
In only a few studies of T4L has pathway identification been quantitatively related to binding kinetics. These studies have employed methods of varying computational efficiency ranging from unbiased MD simulations to the use of machine learning to enhance sampling. A recent example of the latter is reweighted autoencoded variational Bayes for enhanced sampling (RAVE) [17], where a neural network uses information from short unbiased MD simulations to learn progress coordinates. A variation of RAVE combined with the predictive information bottleneck (PIB) [18] framework was developed and applied to benzene unbinding from T4L, resulting in one path (FGH) and a $k_{off}$ 100-fold lower than the experimental value [18,19*]. Weighted ensemble MD (WEMD)[20], which has recently been applied to computing the residence time of a drug with a long τ of 11 minutes [21*], yielded preliminary $k_{off}$ values for benzene unbinding from T4L 100-fold larger than the experimental value [22] (**Table 1**). In contrast, Mondal *et al.* [13] obtained $k_{off}$ values only 3-fold lower than the experimental value using unbiased MD simulations or InMetaD. A similar accuracy was achieved by Wang *et al.* [23] using InMetaD or frequency-adaptive metadynamics (FaMetaD).

The most accurate $k_{off}$ estimation [13] was achieved with sampling of two unbinding paths (FGH, HJ), whereas the less accurate methods sampled one (FGH) [18] or four (FGH, CF, CD, HJ) paths [22]. Indeed, the rates computed for the four paths with WEMD were similar within standard error, indicating that these four paths may be relevant to the binding process. Rydzewski and Valsson also found quite similar unbinding times and average radii for these benzene unbinding paths [10]. Nevertheless, taken together, these results indicate that good $k_{off}$ estimations can be achieved without exhaustive path sampling so long as some of the most probable paths are sampled. Moreover, methods with low computational cost, such as InMetaD and FaMetaD [13,23], are able to provide estimates as accurate as methods of high computational cost, such as unbiased MD simulations.

## Computation of residence times for a highly flexible protein binding site: Heat Shock Protein 90 (N-HSP90) inhibitors.

### *A consistent benchmark dataset of kinetic rate constants measured by SPR for 104 N-HSP90 inhibitors*

Recently, a large set of binding kinetics data measured by surface plasmon resonance (SPR) for drug-like inhibitors of the cancer target N-HSP90 has been made available [24–27]. These compounds bind in the ATP-binding site of the N-terminal domain of the protein (**Figure 2A**). The dataset has more than one hundred compounds that have kinetic rate constants ranging over 4 orders of magnitude for binding to N-HSP90 (**Figure 2B**,[28]). Moreover, for at least 45 of these compounds, there are crystallographic structures of the complexes with N-HSP90 available with a resolution of 2 Å or better. An important characteristic of the N-HSP90 binding site is its flexibility. Depending on the ligand bound, it may adopt various conformations due to the plasticity of the α-helix 3 region, which undergoes different degrees of distortion in the vicinity of residue L107 (see **Figure 2A**). The pure helical conformation is observed solely in a subset of the holo-structures, indicating stabilization of the protein secondary structure by the presence of a ligand in the binding site. Together with the large diversity of compounds in the dataset, the backbone and sidechain conformational variability of the protein binding site pose a significant challenge for the prediction of the dissociation mechanism and (un)binding rates.

## Tradeoff between computational time and sensitivity for ligand unbinding simulations

Several enhanced sampling MD methods for computing $k_{off}$ have been tested on different subsets of the N-HSP90 dataset: τRAMD [24][29*], targeted MD (TMD) [27][30], scaled MD [25] (**Figure 3** and **Table 1**). Apart from one ligand simulated using dissipation-corrected targeted (dcTMD) [30], these studies provide relative values of τ or $k_{off}$, with performance assessed by the correlation between experimental ($\log(k_{off})$) and computed data (either unbinding energy [31] or work [27], or log(unbinding time) [25], [24]; see **Table 1**). In addition, a new complementary approach, CaverDock [32] [33], was applied to a set of 32 N-HSP90 complexes [31]. CaverDock is an adaptation of the Caver approach for detecting ligand egress routes in which ligand dissociation energy profiles are estimated by iteratively docking the ligand along the identified egress routes. Although, it is difficult to compare the performance of the different methods because the data employed in each case only partially overlap, it can be noted that the methods employing longer simulations (on the nanosecond time scale, i.e. [24] and [25]) generally resulted in better accuracy than those with shorter simulations ([27] and [31]). This observation corresponds with the expected tradeoff between the computational time (dependent on the magnitude of the sampling bias that speeds up dissociation) and the sensitivity of the methods to the underlying energy landscape, which is required for the reliable estimation of unbinding rates. As a result of the biases applied, there is incomplete agreement amongst these studies regarding the factors affecting the unbinding rate, the presence of a transition barrier (no barrier found by CaverDock [31] and a clear barrier in TMD [27]), and the dissociation pathways (e.g. the contribution of the alternative exit route to $k_{off}$, **Figure 2C**). Importantly, because of a general correlation between $k_{off}$ and $K_D$ in the N-HSP90 inhibitor data set (**Figure 2B**), even simulations with a large uncertainty in the transition state may lead to a reasonable correlation between computed and experimental $k_{off}$ values.

It is also noteworthy that the best agreement between computed and experimental data was observed for congeneric series of compounds (as highlighted in Refs. [24], [27] and [25]). This hints at cancellation of errors which may relate to deficiencies in the force fields employed, but also to the diversity of the mechanisms affecting τ which are not equally treated by the computational approaches employed.

## Consideration of protein dynamics in enhanced sampling MD simulations of unbinding rates

In all the MD simulation studies of N-HSP90, prior knowledge of the binding site conformation (either loop- or helix-like shape of the α-helix 3 region) was employed. The only attempt to run calculations starting from an apo-structure of N-HSP90 was undertaken with Caverdock [31], but no correlation between computed and measured $k_{off}$ values was found. Moreover, the accuracy of the modeled structures of the protein-ligand complexes was shown to be critical for reliable MD simulations [24][31]. However, even if the crystal structure of the protein-ligand complex is available, additional exploration of the bound state, that cannot be fully sampled in short enhanced dissociation simulations, may be necessary for accurate computation of dissociation energies and unbinding rates. This problem is addressed in the τRAMD[24] and TMD [27] protocols by performing multiple MD simulations to generate an initial equilibrium state ensemble prior to dissociation simulations.
Protein dynamics is treated to different extents in the enhanced sampling methods. For example, restraining backbone motion completely (as in CaverDock [31]) or partially (as in scaled MD [25]) can impede ligand dissociation. Therefore, the latter method was re-designed in selectively scaled MD (ssMD) by scaling only the Lennard-Jones potential between the ligand and water atoms to avoid the necessity to add protein restraints [34]. However, even without additional restraints, it is in general questionable whether biased MD is able to ensure the correct balance between the contributions of the slow and fast degrees of freedom of the system to the ligand unbinding process.

### *QSKR models for N-HSP90 inhibition*

The large dataset of N-HSP90 inhibitors provides a good basis for developing statistical models aimed at the optimization of binding kinetics. Several such models have been reported recently for N-HSP90 – for prediction of τ from protein-ligand interaction fingerprints, PL-IFs, along dissociation trajectories [29] and from binding energy terms for the bound complexes [35], and for prediction of $k_{on}$ from ligand desolvation energies [26]. The first two models are structure-based, which narrows their practical application. However, while a QSKR model dependent only on ligand properties might be preferable, it may require a large dataset and may never achieve the desired accuracy and interpretability. Despite a correlation between binding kinetic parameters and molecular weight being common (larger inhibitors of N-HSP90 tend to have slower binding kinetics than smaller compounds[36]), QSKRs can be expected to be strongly nonlinear since quite diverse factors have been reported to affect τ. In different subsets of the N-HSP90 dataset, these factors have included ligand fragment charge [27], ligand desolvation energy [29] and the degree of protein structural rearrangement required for ligand dissociation [24]. On the other hand, statistical models based on protein-ligand interactions provide an effective way to obtain insight into the molecular determinants and mechanisms underlying binding kinetics. For example, several studies (for N-HSP90 [29], p38 MAP kinase [37], and HIV-1 protease [38]) showed that the unbinding rate was determined by protein-ligand interactions in the first half of the dissociation process, which was also supported by a chemometric study [35] where $k_{off}$, was found to be well predicted from the interactions of the bound state.

## Conclusions and future directions

Taken together, the studies on the T4L and N-HSP90 datasets reviewed here provide a snapshot of the state-of-the-art of molecular simulations to study protein-ligand binding kinetics and their possible applications. The methods are being actively developed (recent examples include ssMD [34], an extension of scaled MD, reweighting of ensembles by variance optimization (REVO) [39] based on WEMD, and FaMetaD [23] a modified metadynamics approach [23]). The methods address a range of needs. For example, a key distinguishing factor between simulation methods is whether they require user-choices about reaction coordinates and if so, which and how complex these choices are. Another important difference is whether absolute or relative values of $k_{off}$ are computed. The optimization of sampling for the computation of absolute values of $k_{off}$ by using different methods, including machine learning, has shown significant progress in reducing simulation time by about an order of magnitude, but these methods are still too computationally demanding for routine use and notably slower than the methods to compute relative $k_{off}$ values (such as τRAMD, TMD and scaled MD). The latter can be applied to a large range of $k_{off}$ values, often with better prediction accuracy, and thus, should be suitable for use in practical applications. Indeed, recently, elABMD, a method based on adiabatic bias MD, and scaled MD have, respectively, been used prospectively to compute τ for a congeneric series of glycogen synthase kinase 3 beta inhibitors [40] and a small set of human D-amino acid oxidase inhibitors [41]. In the next few years, we anticipate further method development and validation studies on a diverse range of relevant drug targets, as well as genuine applications of these methods in drug discovery projects. The success of these studies will however rely on the availability of sizeable (kinetic rate measurements for > 30 compounds), high-quality datasets of protein-ligand binding kinetics.

## Glossary

aMD – Accelerated molecular dynamics [12]
dcTMD – Dissipation-corrected targeted MD [30]
FaMetaD - Frequency-adaptive metadynamics [23]
InMetaD - Infrequent metadynamics [13,23]
MD - Molecular Dynamics simulations [24]
MD+MSM – Unbiased Molecular Dynamics simulations and Markov State Model [13]
MD+AMSM – Multiple MD simulations guided by Adaptive Markov state models (AMSM) [42]
N-HSP90 – N-terminal domain of heat shock protein 90
PIB - Predictive Information Bottleneck, an information theory method to learn a low-dimensional representation of a system that has maximum predictive power [18]
RAMD – Random acceleration molecular dynamics [43]
RAVE - Reweighted autoencoded variational Bayes for enhanced sampling [17,19]
REVO - Reweighting of Ensembles by Variance Optimization [39]
τ – Residence time
τRAMD – A protocol for obtaining relative residence times from RAMD simulations [24,29]
QSKR - Quantitative structure-kinetics relationship
SPR - Surface plasmon resonance
TMD – Targeted MD [27]
T4L – T4 lysozyme
Scaled MD – scaled or smoothed MD in which the systems potential energy is scaled down [25,41]
ssMD - Selectively scaled MD [34]
WEMD – Weighted ensemble molecular dynamics [21,22]

## Acknowledgements

The authors thank Manuel Glaser and Dr. Goutam Mukherjee for suggestions on the manuscript.
Funding: This work was supported by a Capes-Humboldt postdoctoral scholarship to A N-A (Capes process number 88881.162167/2017-01), the European Union's Horizon 2020 Framework Programme for Research and Innovation under Grant Agreement 785907 (Human Brain Project SGA2) and the Klaus Tschira Foundation.

## Declaration of Interest

The authors declare no conflict of interest.

**Table 1:** Methods for computing ligand-receptor dissociation rates illustrated in Figure 3. Methods with recently published applications are listed and the simulation times given correspond to those in these applications.

| Method | System | No. of compl-exes | Range of $k_{off}$ [s$^{-1}$] | Simulation accuracy Pearson correlation coefficient $R^2$ | Error factor (fold) | Computation time per compound [1)] [µs MD] | References |
|---|---|---|---|---|---|---|---|
| PIB | T4L | 1 | $1 \times 10^3$ | n.d. | 100 | ~ 0.1 | [18] |
| FaMetaD | T4L | 2 | $0.3\text{-}1 \times 10^3$ | n.d. | 2-5 | 2-5.5 | [23] |
| InMetaD | T4L | 1 | $0.3 \times 10^3$ | n.d. | 2 | 4.5 | [23] |
| InMetaD | T4L | 1 | $1 \times 10^3$ | n.d. | 3 | n.d. | [13] |
| WEMD | T4L | 1 | $1 \times 10^3$ | n.d. | 1-100 | 29 | [22] |
| MD + MSM | T4L | 1 | $1 \times 10^3$ | n.d. | 3 | 59 | [13] |
| TMD | HSP90 | 25 | $1 \times 10^{-4}\text{-}0.5$ | 0.45-0.8 | n.d. | 0.006 | [27] |
| τRAMD | HSP90 | 70/92 | $1 \times 10^{-4}\text{-}0.5$ | 0.78-0.72 | ~2 | 0.004-0.4 | [24]/ [29] |
| Scaled MD | HSP90 | 7 | $6 \times 10^{-4}\text{-}0.5$ | n.d. | 4-20 | 0.2-7 | [25] |
| dcTMD | HSP90 | 1 | $3 \times 10^{-2}$ | n.d. | 20 | ~1 | [30] |
| Caverdock | HSP90 | 32 | $1 \times 10^{-4}\text{-}0.5$ | 0.63 | n.d. | n.d.[2)] | [31] |
| Combine | HSP90 | 70 | $1 \times 10^{-4}\text{-}0.5$ | 0.8 | ~2 | n.d.[2)] | [35] |

[1]The simulation time depends on the properties of the system studied as well as the methods used. Only production/dissociation simulations are considered, not the system preparation/equilibration.
[2]Methods that are not based on MD simulations.

# Figures and captions

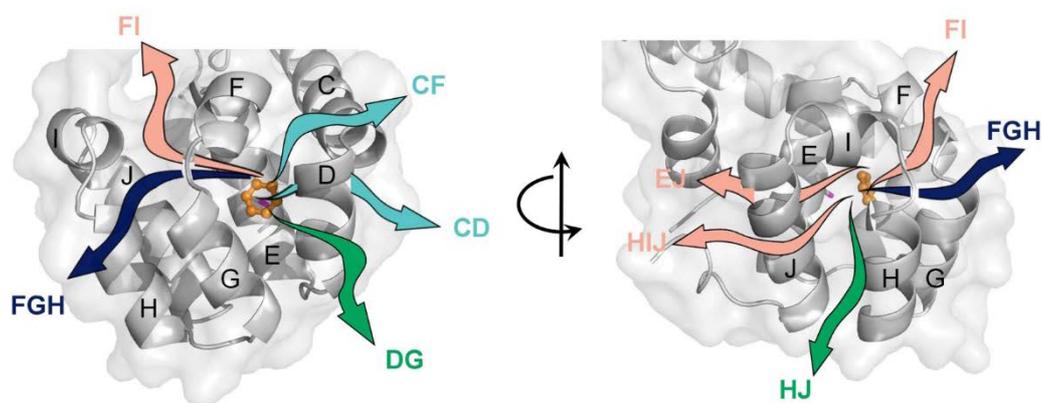

**Figure 1. Exit routes of benzene from the buried cavity in T4L identified in MD simulation studies.** Two views of the structure of the complex (PDB ID 1L83 [44]) of T4L (in gray cartoon with helices labeled with black letters) and benzene (in orange ball and stick representation). The mutated A99 methyl group is shown in pink. Paths for benzene (un)binding are denoted by arrows between the named helices. The arrows are colored according to how often the paths were found in the different studies (from dark blue (most frequent) through green and cyan to salmon (least frequent)): Path FGH: [9,10,12–14,18,19,22,45]; HJ: [9,10,13,22,45]; DG: [9,10,13,14,45]; CF, CD: [9,10,22,45]; FI, HIJ, EJ: [9].

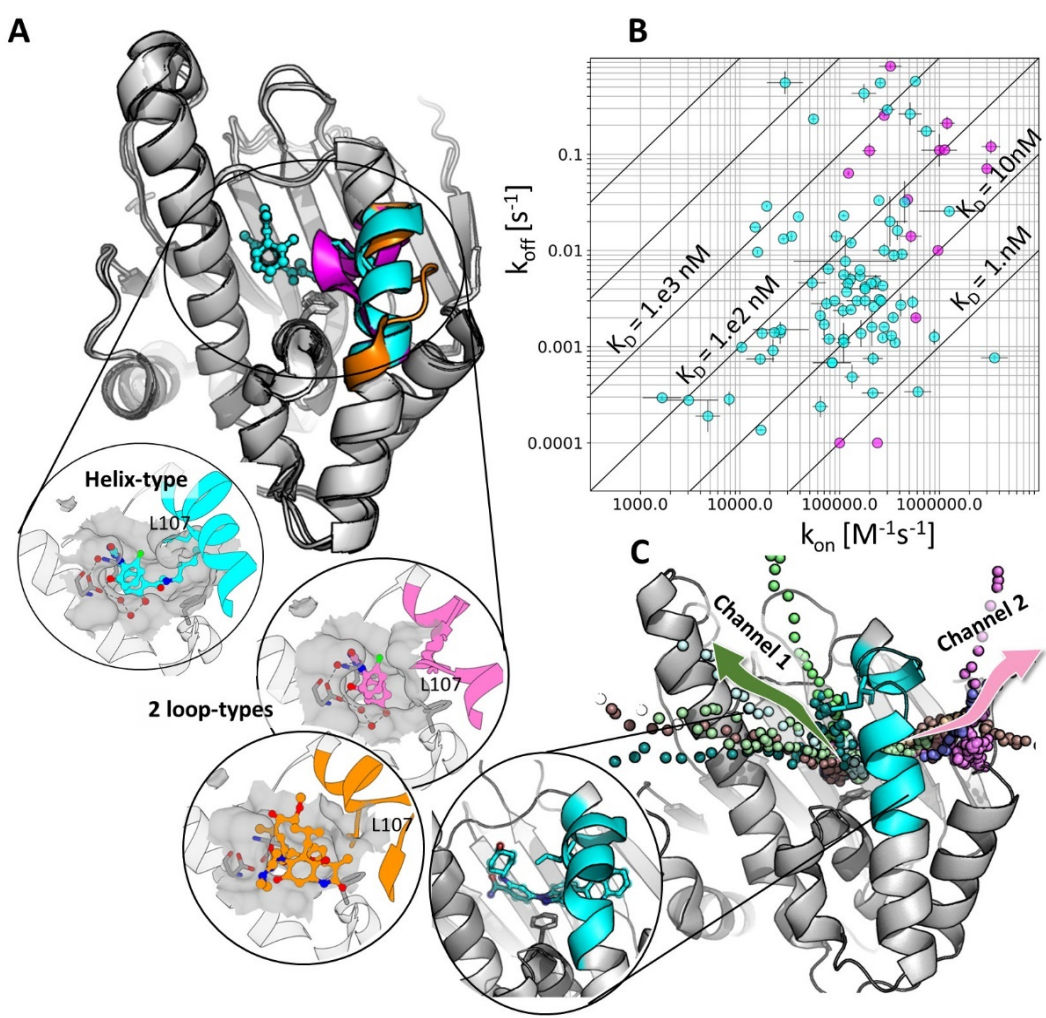

Figure 2: N-HSP90-inhibitor complexes provide a benchmark for the computation of binding kinetics. **(A):** N-HSP90 holo-structures have either helix-type (PDB ID 5J9X [36]) or loop-type (PDB ID 5J64 [46] and 1OSF [47]) conformations of the α-helix 3 segment (the corresponding ligand and flexible part of the α-helix 3 region are highlighted in cyan, magenta, and orange, respectively). **(B):** Plot of $k_{off}$ vs $k_{on}$ for 104 HSP90 inhibitors [28]. Cyan: helix-binders, magenta: loop-binders. **(C):** Representative dissociation routes for the compound shown in the inset (PDB ID 5LQ9[24]) generated in τRAMD simulations are shown by dots colored according to pathway: channel 1 (green) directly from the ATP binding site and channel 2 (pink) through the hydrophobic tunnel under α-helix 3. The latter pathway is observed only for a few compounds that occupy the hydrophobic sub-pocket under α-helix 3 and have a comparatively small moiety in the main ATP binding site (as shown in the inset).

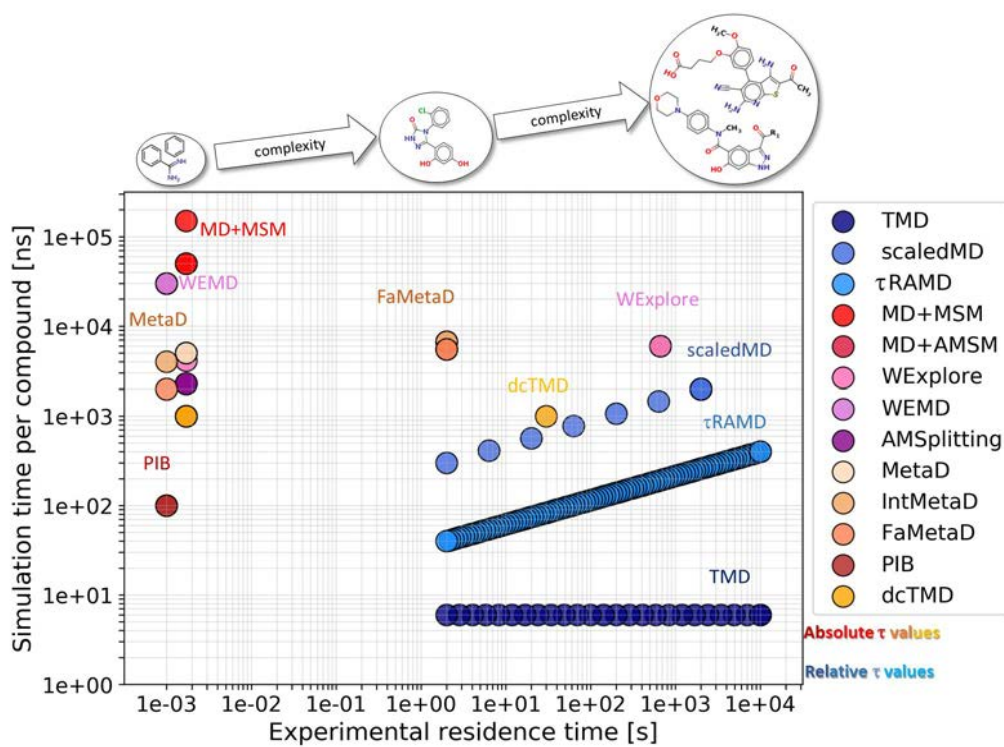

Figure 3: Schematically illustration of the computational performance of MD simulation-based methods for computing protein-ligand dissociation rates. The simulated time required for computing $k_{off}$ is plotted in a log-log plot against the measured residence time for the following systems (with representative ligands shown above the plot): benzene from T4L ($k_{off}$ = $10^3$ s$^{-1}$ [8]), benzamidine from trypsin ($k_{off}$= 600 s$^{-1}$[48]), inhibitors from HSP90 ($k_{off}$ = $10^{-4}$ s$^{-1}$-0.8 s$^{-1}$ ), and the TPPU inhibitor from soluble epoxide hydrolase ($k_{off}$ = 1.5 $10^{-3}$ s$^{-1}$, absolute value computed by WExplore method[21] ); the computational data for T4L and HSP90 are summarized in Table 1, methods are defined in the Glossary. Methods applied to multiple compounds are indicated by a set of points evenly distributed between the minimum and maximum $k_{off}$ values of the dataset used in the simulations. Methods that provide absolute unbinding rates and those that provide relative values are shown in red and blue color pallets, respectively.